\documentclass[11pt,a4paper]{article} 

\usepackage{feynmf,amsmath,amssymb,latexsym,epsfig,floatflt,subfigure}
\def\sst{\scriptscriptstyle}

\def\bea{\begin{eqnarray}}
\def\eea{\end{eqnarray}}

\def\NPB#1#2#3{Nucl. Phys. {\bf B} {\bf#1} (#2) #3}
\def\PLB#1#2#3{Phys. Lett. {\bf B} {\bf#1} (#2) #3}
\def\PRD#1#2#3{Phys. Rev. {\bf D} {\bf#1} (#2) #3}
\def\PRL#1#2#3{Phys. Rev. Lett. {\bf#1} (#2) #3}

\def\NCA#1#2#3{Nuovo Cim. A {\bf#1} (#2) #3}

\def\HEP#1{arXiv:hep-ph/#1}
\def\HEPEX#1{arXiv:hep-exp/#1}
\def\JMP#1#2#3{J. Math. Phys {\bf#1}(#2)#3}
\def\JHEP#1#2#3{JHEP{\bf#1}(#2)#3}
\begin{document} 
\thispagestyle{empty}
\begin{titlepage}

\title{\bf A Scenario for  Spontaneous CP Violation in SUSY SO(10)\footnote{
Based on talks at, Planck06: ``From the Planck Scale to the Electroweak
 Scale'', Paris, May 29 2006 and the Symposium on:``QCD:Facts and Prospects'',
Oberw\"olz, September 10, 2006.}}

\author{Yoav Achiman\footnote{e-mail:achiman@post.tau.ac.il}\\[0.5cm]                   
  School of Physics and Astronomy\\
  Tel Aviv University\\
  69978 Tel Aviv,Israel\\[1.5cm]}

\date{ November 2006}
 
\maketitle
\setlength{\unitlength}{1cm}
\begin{picture}(5,1)(-12.5,-12)
\put(0,0){TAUP 2843/06}
\end{picture}
\parindent 0cm
\begin{abstract}
\noindent
A scenario is suggested for spontaneous $CP$ violation in non-$SUSY$ and
$SUSY\ SO(10)$. The idea is to have a scalar potential which generates 
spontaneously a phase, at the high scale, in the VEV that gives a mass
to the $RH$ neutrinos. The case of the minimal renormalizable $SUSY\ SO(10)$ 
is discussed in 
detail. It is demonstrated that this induces also a phase in the $CKM$ matrix.
It is also pointed out that, in these models, the scales of Baryogenesis, 
Seesaw, Spontaneous $CP$ violation and Spontaneous $U(1)_{PQ}$ breaking are 
all of the same order of magnitude. 
\end{abstract}
\thispagestyle{empty} 
\end{titlepage}
 
\parindent 0pt 


\parindent=0pt
{\Large Introduction}\\

There are three manifestations of CP violation in Nature:\\

1) {\em Fermi scale CP violation} as is observed in the $K$ and $B$ decays
\cite{KB}. This violation is induced predominantly by a complex mixing matrix
 of the quarks ($CKM$).\\

2) {\em The cosmological matter antimatter asymmetry ($BAU$)}
is an indication for high scale $CP$ violation\cite{Sakharov}.
In particular, its most popular explanation via leptogenesis\cite{Yanagida} 
requires $CP$ breaking decays of the heavy right-handed ($RH$) neutrinos.\\

3) {\em The strong $CP$ problem } called also the $QCD$ $\Theta$ problem
\cite{window} lies in the non-observation of $CP$ breaking in the strong
interactions while there is an observed $CP$ violation in the interaction of
quarks.\\

{\em Where is $CP$ violation coming from? Is there one origin to all $CP$ 
breaking phenomena?}\\

It was already suggested \cite{Achiman}\cite{YA} that a spontaneous violation 
of $CP$\cite{T.D.Lee},
at a high scale, via the spontaneous generated phase of the $VEV$ that gives
mass to the $RH$ neutrinos, can be the origin of $CP$ violation.\\

{\Large Why spontaneous violation of $CP$}\quad ?\\

1) It is more elegant and involves less parameters than the intrinsic 
violation in terms of complex Yukawa couplings. The intrinsic breaking 
becomes quite arbitrary in the framework of $SUSY$ and $GUT$ theories.\\

2) Solves the $SUSY$ $CP$ violation problem (too many potentially complex 
parameters) as all parameters are real.\\

3) Solves the strong $CP$ problem at the tree level for the same reason.\\

For good recent discussion of spontaneous $CP$ violation ($SCPV$), with many 
references, see Branco and Mohapatra~\cite{BM}.\\
\pagebreak

{\Large Why $CP$ breaking at a high scale\quad ?}\\

1) Needed to explain the $BAU$. Especially in terms of leptogenesis, i.e.
$CP$ violating decays of heavy neutrinos, it is mandatory.\\

2) $SCPV$ cannot take place in the standard model ($SM$) because of gauge 
invariance. Additional Higgs must be considered and those lead generally to
flavor changing neutral currents ($FCNC$). The best way to avoid these is to 
make the additional scalars heavy~\cite{BM}.\\

3) The scale of $CP$ violation can then be related to the seesaw scale as well
as to the $U(1)_{PQ}$ \cite{PQ} breaking scale, i.e. the ``axion window''
\cite{window}.
\\ 

{\Large The conventional $SO(10)$}\\

Let me start by revising the {\em renormalizable non-$SUSY$ $SO(10)$} 
and a possible $SCPV$~\cite{YA}. 
Conventional $SO(10)$  requires  intermediate  gauge symmetry 
breaking ($I_i$)\cite{inter} to have gauge coupling unification.
$$
SO(10) \longrightarrow I_i \longrightarrow SM = SU_C(3)\times
SU_L(2)\times U_Y(1)\ .
$$
Most  models involve an intermediate scale at $\approx 
10^{12} GeV$ for:\\
$\bullet$ Breaking of $B-L$ \\ 
$\bullet$ The masses of $RH$ neutrinos\\
$\bullet$  $CP$ violation~responsible~for~
leptogenesis ($BAU$)\\

$SO(10)$ fermions are in three ${\bf 16}$ representations: $\Psi_i(\bf 16)$.
$$
{\bf 16}\times{\bf 16} = ({\bf 10} + {\bf 126})_S
+ {\bf 120}_{AS}\ .
$$

Hence, only $H({\bf 10)},\ {\overline\Sigma}(\overline{\bf 126})$\ and\
$D({\bf 120})$ can contribute directly to Yukawa couplings and 
fermion masses. Additional Higgs representations are needed for the 
gauge symmetry breaking. \\
One and only one $VEV$ \quad
${\overline\Delta} = <{\overline\Sigma} (1,1,0) >$\quad
can give a (large) mass to the $RH$ neutrinos via
$$
Y_\ell ^{ij} \nu_{\scriptscriptstyle{R}}^i {\overline\Delta}
 \nu_{\scriptscriptstyle{R}}^j 
$$
and so induces the seesaw mechanism.
It breaks also \quad {$B-L$}\quad and $SO(10)\rightarrow 
SU(5)$.\\
\pagebreak

{\Large Spontaneous $CP$ violation in conventional $SO(10)$}\\

${\overline\Sigma}({\overline {126}})$ is the only relevant complex Higgs 
representation. Its other special property is that $({\overline\Sigma})_
{\sst{S}}^{\sst{4}}$
is invariant in $SO(10)$~\cite{HHR}. This allows for a $SCPV$ at the high
scale, using the scalar potential:~\cite{YA}
$$
V = V_0 + \lambda_1(H)_{\sst{S}}^{\sst{2}} [({\overline\Sigma})_
{\sst{S}}^{\sst{2} }
+ ({\overline\Sigma}^*)_{\sst{S}}^{\sst{2}}] +
\lambda_2[({\overline\Sigma})_{\sst{S}}^{\sst{4}} + 
({\overline\Sigma}^*)_{\sst{S}}^{\sst{4}}]\ .
$$
Inserting the $VEV$s

$$
<H(1,2,-1/2)> = \frac{v}{\sqrt{\sst{2}}}
\ \ \ \ \  \overline\Delta = \frac{\sigma}{\sqrt{\sst{2}}}
{e^{i\alpha}}
$$

in the neutral components, the scalar potential reads
$$
V(v,\sigma,\alpha) = A\cos (2\alpha) + B\cos (4\alpha)\ .
$$
For $B$ positive and $|A| > 4B$ the absolute minimum of the 
potential requires
$$
\alpha = \frac{1}{2}\arccos\left( \frac{A}{4B}\right ). 
$$
This ensures  the spontaneous breaking of $CP$ \cite{branco}.\\

However, $\Phi^4$ cannot be generated from the superpotential 
 in renormalizable $SUSY$ theories and a different approach is needed
there.\\

{\Large Renormalizable $SUSY SO(10)$ models }\\

Became very popular recently~\cite{maryland}~\cite{trieste}~\cite{japan}~
\cite{wuppertal}
due to their simplicity, predictability and automatic $R$-parity invariance
(i.e. a dark matter candidate).\\

I will limit myself here to the so called {\em minimal model} \cite{minimal}.

It involves the following Higgs representations
$$
H({\bf 10}),\quad \Phi ({\bf 210}), \quad\Sigma ({\bf 126}) \oplus {\overline
\Sigma} ({\overline{\bf 126}})\ .
$$
Both $\Sigma$ and $\overline{\Sigma}$ are required to avoid high scale $SUSY$ 
breaking ($D$-flatness) and $\Phi ({\bf 210})$ needed for the gauge breaking.
 \\

The properties of the model are dictated by the superpotential. This involves
all possible renormalizable products of the superfields

$$
W=M_\Phi \Phi^2 + \lambda_\Phi \Phi^3 + M_\Sigma \Sigma\overline\Sigma
+ \lambda_\Sigma \Phi\Sigma\overline\Sigma
$$
$$
+ M_{\scriptscriptstyle H} H^2 + \Phi H(\kappa\Sigma + \bar\kappa
\overline\Sigma) + 
\Psi_i(Y_{\scriptscriptstyle\bf 10}^{ij}H 
+ Y_{\overline{\scriptscriptstyle\bf 126}}^{ij} \overline\Sigma)\Psi_j
$$
(One can, however, add discrete symmetries or $U(1)_{PQ}$ etc. on top of
$SO(10)$.)\\

We take all coupling constants real and positive, also in the soft $SUSY$
breaking terms.\\
\pagebreak

The symmetry breaking goes in two steps

$$
SUSY SO(10) \stackrel{strong\ gauge\ breaking}
{\longrightarrow} MSSM 
\stackrel{{\scriptscriptstyle SUSY}\ breaking}{\longrightarrow} SM
$$

The $F$ and $D$-terms must vanish during the strong gauge 
breaking to avoid high scale $SUSY$ breakdown ("$F$,$D$ 
flatness").\\

{\em $D$-flatness}:\quad
 only $\Sigma$, $\overline\Sigma$ are relervant therefore
$$
|\Delta| = |\bar\Delta| \ \hbox{ i.e. }\ \sigma  = \bar\sigma\ .
$$ 

The situation with\quad {\em $F$-flatness}\quad is more complicated.\\ 
The strong breaking
 is dictated by the $VEV$s that are $SM$ singlets.\\ 
Those are in the ${\scriptstyle SU_C(4)\times SU_L(2) \times SU_R(2})$  
notation :

$$
\phi_1 = <\Phi(1,1,1)> \ \  \phi_2 = <\Phi(15,1,1)> \ \  
\phi_3 = <\Phi(15,1,3)>
$$ 
$$
\Delta = <\Sigma(\overline{10},1,3)> \ \ \ \  \bar\Delta = 
<\bar\Sigma(10,1,3)> .
$$
 
The strong breaking superpotential in terms of those $VEV$'s is then
$$
\begin{array}{ccl}
W_H & = & M_\phi(\phi_1^2 + 3\phi_2^2 + 6\phi_3^2)
   + 2\lambda_\phi(\phi_1^3 + 3\phi_1\phi_2^2 + 6\phi_2\phi_3^2)\\
[10pt]
&+& M_W\Delta\bar\Delta + \lambda_\Sigma\Delta\bar\Delta
(\phi_1 +3\phi_2 + 6\phi_3) .
\end{array}
$$

$\sst \left| \frac{\partial W_H}{\partial 
v_i}\right|^2 = 0$ \qquad
gives a set of equations.
Their solutions dictate the details of the strong symmetry breaking.
One chooses the parameters such that the breaking 
$$
SUSY SO(10) \longrightarrow MSSM
$$
will be achieved~\cite{Senjan}~\cite{Fukuyama}.\\

$SUSY$ is broken by the soft $SUSY$ breaking terms.
The gauge $MSSM$ breaking is induced by the $VEV$'s
of the $SM$ doublet $\phi^{u,d}(1,2,\pm 1/2)$ components of the Higgs 
representations.\\

The mass matrices of the Higgs are then as follows
$$
M_{ij}^u = \left[\frac{\partial^2W}{\partial\phi_i^u
\partial\phi_j^u}\right]_{\phi_i=<\phi_i>}\ \ \ 
M_{ij}^d = \left[ \frac{\partial^2W}{\partial\phi_i^d
\partial\phi_j^d}\right]_{\phi_i = <\phi_i>} .
$$
The requirement 
$$
\det(M_{ij}^u) \approx 0 \ \ \ \det(M_{ij}^d)\approx 0
$$
leaves only two light combinations of doublet components and those play the 
role of the bidoublets \quad $h_u , h_d$ \quad of the $MSSM$. (This also is 
discussed in detail in the papers of~\cite{Senjan}~\cite{Fukuyama}.) \\

We will come back to $h_u , h_d$ later but let me discuss the $SCPV$ 
first.\\

\pagebreak
{\Large Spontaneous $CP$ violation in $SUSY$ SO(10)}\\

As in the non-SUSY case, we conjecture that $\Delta$ and $\bar\Delta$, 
and only those, acquire a phase at the tree level
$$
<\Sigma(1,1,0)> \equiv \Delta = \sigma e^{i\alpha}\ \ \ \ 
<\bar\Sigma(1,1,0)> \equiv \bar\Delta = \sigma e^{i\bar\alpha} .
$$

Let me show that this is a minimum of the scalar potential in a
certain region of the parameter space.


To do this we collect all terms with 
$\Delta, \bar\Delta$ in the 
superpotential. Those involve the $VEV$'s that are non-singlets 
under the $SM$. I.e. the $SM$ doublet components of the Higgs 
representations.
$$ 
\begin{array}{ccccccc}
\phi^u&=&<\Phi(1,2,1/2)>&\ \ \ &\phi^d&=&<\Phi(1,2,-1/2)>\\
H^u &=& <H(1,2,1/2)>&\ \ \ &H^d &=& <H(1,2,-1/2)>\\
\Delta^u&=&<\Sigma(1,2,1/2)>& & \Delta^d&=&<\Sigma(1,2,-1/2)>\\
\bar\Delta^u&=&<\bar\Sigma(1,2,1/2)>& &\bar\Delta^d & = &
<\bar\Sigma(1,2,-1/2)>\\
\end{array}
$$
The relevant terms are:
\begin{eqnarray*}
W_\Delta &=& M_\Sigma \Delta\bar\Delta + 
\frac{\lambda_\Sigma}{10} (\phi^u\Delta^d\bar\Delta +
\phi^d\bar\Delta^u\Delta)\\
&+& (\frac{\lambda_\Sigma}{10}(\frac{1}{\sqrt{6}}\phi_1\Delta
\bar\Delta + \frac{1}{\sqrt{2}}\phi_2\Delta\bar\Delta
+\phi_3\Delta\bar\Delta )\\
&+& \frac{\lambda_\Sigma\sqrt{2}}{15} \phi_2\bar\Delta^u
\Delta^d - \frac{\kappa}{\sqrt{5}} \phi^d H^u\Delta -
\frac{\bar\kappa}{\sqrt{5}} \phi^u H^d \bar\Delta
\end{eqnarray*}
using ~\cite{Senjan}~\cite{Fukuyama}.\\ 

One can then calculate the corresponding scalar potential
$$
V(\alpha,\bar\alpha,M_\Sigma,\lambda_\Sigma,\kappa,\bar\kappa,v_i) =
\sum_i \left| \frac{\partial W_\Sigma}{\partial v_i}
\right|^2\ .
$$
Noting that  \qquad$|A+Be^{i\alpha} |^2 = A^2 + B^2 +2AB \cos\alpha$\\

and \qquad $|K+P\Delta\bar\Delta|^2 = K^2 + P^2\sigma\bar\sigma +
2KP\sigma\bar\sigma\cos(\alpha+\bar\alpha)$,\\

one finds that\\

$V=A(M_\Sigma, \lambda_\Sigma, \kappa, \bar\kappa, v_i) + 
B(M_\Sigma, \lambda_\Sigma, \kappa, \bar\kappa, v_i)\cos\alpha+$\\

$ D(M_\Sigma, \lambda_\Sigma, \kappa, \bar\kappa ,v_i)
\cos\bar\alpha + E(M_\Sigma, \lambda_\Sigma, \kappa, \bar\kappa,
v_i)\cos(\alpha+\nobreak\bar\alpha)$ .\\

For explicit expressions of the coefficients see the Appendix.\\

The minimalization under $\alpha, \bar\alpha$ requires

$$
\begin{array}{ccc}  
\frac{\partial V(\alpha)}{\partial\alpha}&=& -B\sin\alpha
- E\sin(\alpha+\bar\alpha) = 0\\[5pt]
\frac{\partial V(\bar\alpha)}{\partial\bar\alpha} 
&=& -D\sin\bar\alpha - E\sin(\alpha + \bar\alpha) = 0 
\end{array} \ .
$$\\

\pagebreak

This gives the equations

$$
\sin\bar\alpha = 
\frac{B}{D}\sin\alpha\ \ \ \ 
$$ 
$$
B\sin\alpha+E(\sin\alpha\cos\bar\alpha+\sin\bar\alpha\cos\alpha)=0
$$


and the solutions are

$$
\cos\alpha=\frac{ED}{2}(\frac{1}{B^2}-\frac{1}{D^2}-\frac{1}{E^2})
$$
$$
\cos\bar\alpha=\frac{EB}{2}(\frac{1}{D^2}-\frac{1}{B^2}-\frac{1}{E^2}) .
$$
\\
We have clearly a minimum for a certain range of parameters, with non trivial
values of $\alpha$, $\bar\alpha$. This means that $CP$ is broken 
spontaneously.\\

{\Large Our $SCPV$ induces a phase in the $CKM$ matrix}\\

We mentioned already that the $MSSM$ bi-doublets $h_u,h_d$ are
given (linear) combinations 
of the Higgs representations doublet components. 
The general explicit combination are given in~\cite{Senjan} and partially 
also in~\cite{Fukuyama}. Those expressions are quite complicate so let me skip 
them and refer you to the above papers.\\ 
The important relevent  fact for us is that coefficients of those 
combinations involve $\Delta$ and $\bar\Delta$ (and a possibly complex
parameter $x$  that fixes the local symmetry breaking \cite{Senjan}) 
so that the $VEV$s\quad $<h_u>$,$<h_d>$ \quad are complex. \\

$H$ and $\overline\Sigma$ which come in the Yukawa coupling and contribute
to the mass matrices
$$
M^i=Y_{10}^iH + Y_{\overline {126}}^i\overline\Sigma\\
$$
are given in 
terms of the physical $h_{u,d}$ as follows (the heavy combinations decouple):
$$
\begin{array}{cccccccc}  
H_{u,d}&=&a_u h_u &+& a_d h_d &+&\cdots&{\hbox{\small decoupled}}\\\
\bar\Sigma_{u,d}&=&b_u h_u& +& b_d h_d &+&\cdots& 
{\hbox{\small decoupled}}\\
\end{array}
$$
\\
The mass matrices are expressed then in terms of $<h_{u,d}>$
$$
M_u=(a_uY_{10}+b_uY_{\overline{126}})<h_u>
$$
$$
M_d=(a_dY_{10}+b_dY_{\overline{126}})<h_d>
$$
$$
M_\ell=(a_dY_{10}-3b_dY_{\overline{126}})<h_d>
$$
$$
M_{\nu}^D=(a_uY_{10}-3b_uY_{\overline{126}})<h_u>
$$
$$
M_{\nu_R}=Y_{\overline{126}}\bar\Delta 
$$
\\

The mass matrices of the quarks and also leptons are therefore complex and
lead to a complex $CKM$ matrix as well as a complex $PNMS$ leptonic one.\\ 

\pagebreak

{\Large Remarks concerning other $SCPV$ models}\\

To the best of my knowledge there are no $SUSY\ GUT$ models that really
discuss the way the phases are generated spontaneously. $SCPV$ is
induced in most models in giving adhoc phases by hand to some of the $VEV$s.\\
 
{\Large Is the $SCPV$ related to the strong $CP$ problem?}\\

The spontaneous breaking of $CP$ solves the $QCD$ $\Theta$ problem but only 
at the tree
level. To suppress also radiative corrections, {\em a la} Barr~\cite{Barr} and
Nelson~\cite{Nelson},  one must however go beyond $SO(10)$. The simplest
solution, in the framework of the renormalizable $SO(10)$, is to require
global $U(1)_{PQ}$~\cite{PQ} invariance with the invisible axion scenario~\cite{axion}. It is interesting then to observe that the energy range of our $SCPV$ lies within the invisible axion window~\cite{window}
$$
10^9 GeV \stackrel{<}{\sim} f_a \stackrel{<}{\sim}
10^{12} GeV\ ,
$$
where $f_a$ is the axion decay constant.\\

This can  be applied to $SUSY SO(10)$ as well.
The minimal renormalizable $SUSY SO(10)\times U(1)_{PQ}$ was discussed 
recently in a paper by Fukuyama and Kikuchi~\cite{F+K}. The requirement
of $U(1)_{PQ}$ invariance using the $PQ$ charges
$$
PQ(\Psi)=-1,\quad PQ(H)=2, 
$$
$$
PQ(\Sigma)=-2,\quad PQ(\overline\Sigma)=2,\quad PQ(\Phi)=0
$$
forbids only two terms in the superpotential
$$
W_{PQ}=M\Phi^2 + \lambda_\Phi \Phi^3 + M_\Sigma \Sigma\overline\Sigma
+ \lambda_\Sigma \Phi\Sigma\overline\Sigma
$$
$$
 + {\cal K}\Phi\Sigma H + 
\Psi_i(Y_{\scriptscriptstyle\bf 10}^{ij}H 
+ Y_{\overline{\scriptscriptstyle\bf 126}}^{ij} \overline\Sigma)\Psi_j\quad . 
$$ 
Hence, our scenario for $SCPV$ is still intact (although with different 
phases).\\

The breaking of local 
$B-L$ via the $VEV$s of $\overline\Sigma(\overline{126})$ and 
$\Sigma(126)$ will  
also break spontaneously the global $U(1)_{PQ}$ and explain the coincidence of 
the scales of the axion window and the seesaw one. In our scenario it will
also coincide with the scale of $SCPV$ and that of leptogenesis.\\

Fukuyama and Kikuchi~\cite{F+K} suggest in their paper that the difference
between the phases of $\Delta$ and $\bar\Delta$ is related to the
axion\footnote{G.Senjanovic claims however that it is not possible to break two symmetries using one $VEV$ (private communication after my talk in Paris).}.
\\

\pagebreak


{\Large Conclusions}\\

I presented, in these talks, a scenario for $SCPV$ in both non-$SUSY$ and $
SUSY \ SO(10)$.
$CP$ is broken spontaneously at the scale of the $RH$ neutrinos but a phase
is generated also in the $CKM$ low energy mixing matrix. We have therefore
$CP$ violation at low and high energies as is required experimentally.\\
If $U(1)_{PQ}$ invariance is also used, one finds the interesting situation
that the scales of Baryogenesis, Seesaw, $SCPV$ and the breaking of 
$U(1)_{PQ}$ are all at the same order of magnitude.\\ 
A detailed paper based on the above talks is in preparation.\\

{\Large Appendix: the parameters of the scalar potential}\\

$$
\frac{\partial W_\Delta}{\partial \phi^u},\ \   \frac{\partial W_\Delta}
{\partial \phi^d},\ \   \frac{\partial W_\Delta}{\partial \phi_1},\ \
\frac{\partial W_\Delta}{\partial \phi_3},\ \   \frac{\partial W_\Delta}
{\partial H^u},\ \   \frac{\partial W_\Delta}{\partial H^d}
$$
do not give terms with a phase.\\

$\alpha$ dependent terms are obtained from\quad $ \frac{\partial W_\Delta}
{\partial \bar\Delta}$\quad and \quad $\frac{\partial W_\Delta}
{\partial \bar\Delta^u}$ 
\quad i.e.\\
$$
\left|\frac{\partial W_\Delta}{\partial \bar\Delta}\right|^2+
\left|\frac{\partial W_\Delta}{\partial \bar\Delta^u}\right|^2=constant +
B\cos\alpha
$$ .

Therefore,
$$
B=2\sigma\phi^u[M_\Sigma
+\frac{\lambda_\Sigma}{10}(\frac{1}{\sqrt6}\phi_1+\frac{1}{\sqrt2}\phi_2 +
\phi_3)][\frac{\lambda_\Sigma}{10}\Delta^d - \frac{\bar\kappa}{\sqrt5}H^d]
+\frac{\sqrt2}{75}\sigma\lambda_\Sigma^2\phi^d\phi_2\Delta^d=
$$
$B(M_\Sigma, \lambda_\Sigma, \bar\kappa, \phi_i,\phi^u,\phi^d,
\Delta^d,H^d )$ .\\

In the same way
$$
D=2\sigma\phi^d[M_\Sigma
+\frac{\lambda_\Sigma}{10}(\frac{1}{\sqrt6}\phi_1+\frac{1}{\sqrt2}\phi_2 +
\phi_3)][\frac{\lambda_\Sigma}{10}\Delta^u - \frac{\bar\kappa}{\sqrt5}H^u]
+\frac{\sqrt2}{75}\sigma\lambda_\Sigma^2\phi^u\phi_2\Delta^u=$$
$D(M_\Sigma, \lambda_\Sigma,\kappa, \phi_i,\phi^u,\phi^d,
\Delta^u,H^u) $ .\\

A term proportional to\quad $\cos(\alpha+\bar\alpha)$\quad is generated only by
$\frac{\partial W_\Delta}{\partial \phi_2}$.\\
Hence,
$$
E=\frac{1}{75}\lambda_\Sigma^2\bar\Delta^u\Delta^d\sigma^2=
E(\lambda_\Sigma,\sigma,\bar\Delta^u,\Delta^d)\ .
$$\\

\pagebreak


\end{document}